# Cosmology at a Crossroads: Tension With the Hubble Constant

**Wendy L. Freedman**

We are at an interesting juncture in cosmology. With new methods and technology, the accuracy in measurement of the Hubble constant has vastly improved, but a recent tension has arisen that is either signaling new physics or as-yet unrecognized uncertainties.

Just under a century ago, Edwin Hubble revolutionized cosmology with his discovery that the universe is expanding. Hubble found a relationship between radial velocity and the distance to nearby galaxies, determining the proportionality constant $H_o$ (=v/r), that now bears his name. The Hubble constant remains one of the most important parameters in cosmology. An accurate value of $H_o$ can provide a powerful constraint on the cosmological model describing the evolution of the universe. In addition, it characterizes the expansion rate of the Universe at the current time, defines the observable size of the Universe, and its inverse sets the expansion age of the Universe.

Hubble (1929) originally measured a value of $H_o$ = 500 km $s^{-1}$ $Mpc^{-1}$. Later revisions led to a range between 50 and 100. Resolution of this discrepancy ultimately required the ability to measure accurate distances: a new generation of digital detectors and the launch of the Hubble Space Telescope (HST). As part of the Hubble Key Project Freedman et al. (2001) measured the value of $H_o$ to be 72 ±2 (statistical) ±7 (systematic) km $s^{-1}$ $Mpc^{-1}$. Since that time, the error bars on $H_o$ have continued to decrease, but this locally and directly measured value of $H_o$ has remained largely unchanged.

Over the past 15 years, measurements of the fluctuations in the temperature of the remnant radiation from the Big Bang have provided a relatively new means of estimating the value of the Hubble constant. This very different approach has led us to an interesting crossroads, yielding a lower derived value of $H_o$ (see Figure 1). If this discrepancy persists in the face of newer and higher precision and accuracy data, it may be signaling that there is new physics to be discovered beyond the current standard model of cosmology.

The classical (local) route to an accurate value of $H_o$ currently has two distinct components: (1) the calibration of stellar luminosities and distances to nearby galaxies (traditionally using Cepheid variables), and (2) the calibration of more luminous objects (Type Ia supernovae) providing distances extending well into the distant smooth Hubble flow. The prescription for measuring accurate Cepheid distances to galaxies has been well-tested and is described in detail elsewhere (e.g., Madore & Freedman 1991; Freedman et al. 2001; Riess et al. 2016). Briefly, Cepheids are identified at optical wavelengths (where the amplitude of variability is largest), and then followed up at longer wavelengths where corrections for extinction due to dust are minimized. The Cepheid distance scale is now anchored to geometric parallax measurements of stars within the Milky Way (and tested using several independent techniques). In the second step, relative distances to galaxies out to cosmological redshifts of $z \sim 0.1$ ($v = 30,000$ km s$^{-1}$) are measured from the peak brightnesses of Type Ia supernovae (SNe Ia). Cepheid variables are identified in nearby galaxies that have well-observed SNe Ia to provide the absolute calibration needed for the determination of $H_o$.

Since the time of the Hubble Key Project, there has been tremendous progress in decreasing known systematic errors. The improvements have come with increases to the samples of SNe~Ia galaxies (Riess et al. 2016), a geometric measurement of the distance to the nearby galaxy NGC 4258 (Humphreys et al. 2013), geometric parallaxes for a sample of Milky Way Cepheids (Benedict et al. 2007); and new *Spitzer* mid-infrared measurements of Cepheids in the Milky Way and the Large Magellanic Cloud (Freedman et al. 2012). All of these refinements yield values of $H_o$ consistent with 73 km s$^{-1}$ Mpc$^{-1}$ to within a margin of error of a few percent. The most recent Riess et al. value asserts an uncertainty of only 2.4%. Independently, measurements of time delays for 2 new gravitational lens systems by the H0LiCOW survey (Bonvin et al. 2017) lead to a consistent value of 71.9 ± 2.7 (±3.8%) km s$^{-1}$ Mpc$^{-1}$. With increasing numbers of lenses in future, this method can, in principle, yield an accuracy of 1%.

The European Space Agency's Planck satellite has recently acquired the highest-sensitivity and highest-resolution maps to date of the sky in microwaves, providing a snapshot of the early universe about 380,000 years after the Big Bang. An analysis of variations in the temperature and polarization maps leads to a striking agreement with the current standard model of cosmology (Planck Collaboration 2016). Fitting the angular power spectrum of fluctuations in the Planck data to a 6-parameter ΛCDM model leads to a derived value of $H_o$ of 67.8 ± 0.9 (±1.3%) km s$^{-1}$ Mpc$^{-1}$, which is over 3-σ discrepant with the most recent Riess et al. (2016) value. The Planck value agrees well with the value of 67.3 ± 1.1 obtained from measurements of baryon acoustic oscillations in combination with SNe Ia (Aubourg et al. 2015).

It is certainly worth noting that the local measurement of Ho is based on the astrophysics of stars, and the CMB results are based on the physics of the early universe: the results are entirely independent of each other. 13.8 billion years of evolution of the universe has occurred since the surface of last scattering of the CMB and the present day, and yet the two measures agree to within 10%. Viewed from a historical perspective, the agreement is actually rather remarkable.

However, the currently estimated error bars do not overlap. Is the discrepancy real or is this a 'tension in a teapot'? The obvious possibility is that one or both of the methods may suffer from unknown systematic errors. In the case of the local distance scale, it is now necessary to rule out systematic errors at the percent level in the Cepheid calibration. Systematic effects may include metallicity variations and photometry biases in high-density crowded regions. However, current, independent tests of the Cepheid distance scale suggest that it is robust to within a few percent (Hatt et al. 2017); the current 8% difference between the two methods is challenging to reconcile. With regard to CMB modeling, currently observed discrepancies between measurements at low $l$ (large spatial scales) and high $l$ (small spatial scales) are not yet completely understood. The more interesting possibility is that there is physics beyond the current standard model; possibilities include decaying dark matter, evolving dark energy, dark radiation, modified gravity or deviations from flatness. For example, additional radiation beyond the standard model, perhaps an additional neutrino or other relativistic species, would have the effect of increasing the expansion rate at early times, and could explain the discrepancy. Yet the analysis of the Planck Collaboration (2016) does not favor any of these possibilities.

To break the current impasse, the steps in the extragalactic distance scale will need to be tested at the percent level. This is a tall order, but new developments are likely to provide a definitive resolution well within the next decade. Closest to home, the European satellite, Gaia, is poised to revolutionize the foundation of the distance scale. The calibration of the extragalactic distance scale will be established at greater than 1% accuracy via geometric parallax measurements of Cepheids, RR Lyrae stars, and red giant branch stars in the Milky Way. These observations will provide an unprecedented improvement in the measurement of $H_o$, finally eliminating the long-standing challenges associated with the absolute zero-point calibration of the extragalactic distance scale.

In parallel, a major new development in the stellar distance scale will come with measurements completely independent of the Cepheid distance scale. Using HST, we have been strengthening the astrophysical distance scale – that based on well-understood and well-studied classes of stars. The most promising complementary route to the Cepheids is one that uses the brightest red giant branch stars in galaxies (see Beaton et al 2016 and references therein). The luminosities of these stars are set by standard nuclear astrophysics. At the transition from their hydrogen-shell burning phase, the temperature in the degenerate, helium cores of these stars exceeds $10^8$ K, provoking a thermal runaway that lifts the pressure-supported degeneracy. The stars terminate their ascent of the red giant branch, and rapidly decrease in luminosity as they begin stable helium burning. This termination of the red giant branch (or TRGB) provides a luminous, easily measured and calibrated method for distance determination (see Figure 2). All galaxies contain a large population of red giants (whereas Cepheid variables occur in spiral galaxies only). For nearby galaxies, it is competitive in terms of precision and accuracy with Cepheids. In addition, these stars are found in the halos of galaxies, avoiding the dust and crowded regions in the spiral arms that are home to the Cepheids. We are currently monitoring more than 1,000 of these giant stars in the Milky Way. With multicolor photometry and Gaia parallaxes, and eventually an extension of the distance scale using JWST, the TRGB will provide a calibration of SNe Ia to better than 1% precision and accuracy, completely independent of the Cepheid distance scale.

The history of cosmology has abundant examples both of discrepancies that ushered in new discoveries, and others that turned out to be unknown systematic errors. Based on the current data, I believe that the jury is still out. But the upcoming results from Gaia, the future availability of JWST, as well as results from upcoming CMB experiments (the South Pole Telescope, SPT and the Atacama Cosmology Telescope, ACT) make the prospects bright for definitively settling the issue in the near future.

*Wendy L. Freedman is the John and Marion Sullivan University Professor of Astrophysics at the University of Chicago, 5640 Ellis Ave, Chicago IL, 60637. e-mail: wlfreedman@uchicago.edu*

**Figures:**

Figure 1: The Current Tension in the Determination of H₀

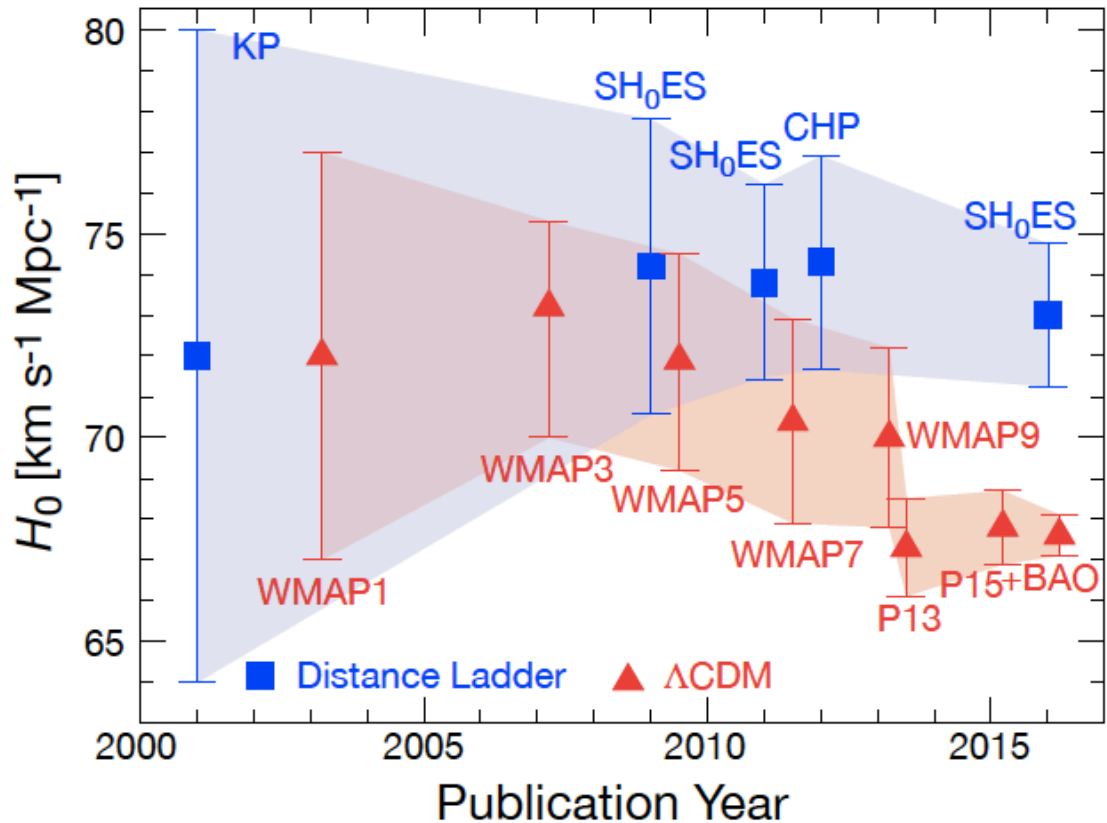

Figure 1: Recent values of $H_o$ as a function of publication date since the Hubble Key Project (adapted from Beaton et al. 2016). Symbols in blue represent values of $H_o$ determined in the nearby universe with a calibration based on the Cepheid distance scale. Symbols in red represent derived values of $H_o$ based on an adopted cosmological model and measurements of the CMB. The blue and red shaded regions show the evolution of the uncertainties in these values, which have been decreasing for both methods. The most recent measurements disagree at greater than 3-σ.

Figure 2: The Tip of the Red Giant Branch (TRGB) For Measuring Distances

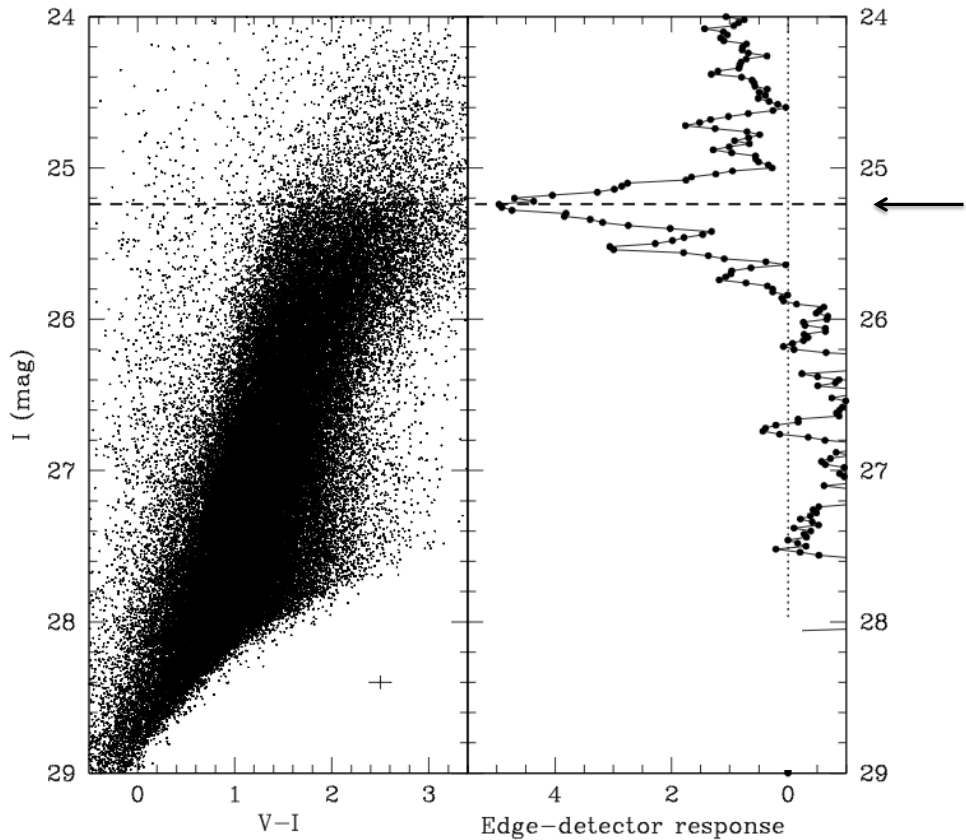

Figure 2: .

Left Panel: An *I-(V-I)* color-magnitude diagram for red giant branch stars in the halo of the spiral galaxy NGC 4258, also host to a $H_2O$ megamaser. The horizontal dashed line indicates the position of the tip of the red giant branch (TRGB). The tip position is measured using a Sobel edge-detection filter.

Right panel: Response of the edge-detection filter. The distance to a galaxy follows simply from the measured the apparent magnitude of the TRGB and its absolute calibration. The absolute calibration of the tip of the red giant branch (currently measured to be at an *I*-band magnitude of -4 mag) will ultimately be calibrated based on geometric parallaxes of Milky Way red giants measured using Gaia.

Reproduced from Mager et al. 2008, ApJ, 689, 721.